\documentclass[usenatbib]{mn2e}
\usepackage{amsmath,amssymb,amsfonts,bm}
\usepackage{graphicx}
\usepackage{hyperref}
\usepackage{calc}
\usepackage{color}
\usepackage{wrapfig}
\usepackage{url}
\usepackage{paralist}
\usepackage{array}
\usepackage{tabularx}
\usepackage{txfonts}
\def\msun{M_\odot}
\def\Msun{M_\odot}

\title{An analytic model for non-spherical lenses in covariant MOdified Newtonian Dynamics}
\author[HuanYuan Shan, Martin Feix, Benoit Famaey and HongSheng Zhao]{HuanYuan Shan$^{1}$\thanks{E-mail:
shanhuany@bao.ac.cn (HYS); mf256@st-andrews.ac.uk (MF)}, Martin Feix$^{2}$\footnotemark[1], Benoit Famaey$^{3}$ and HongSheng Zhao$^{2,1}$\\
$^{1}$National Astronomical Observatories, Chinese Academy of Sciences, 20A Datun Road, Chaoyang District, 100012, Beijing, China\\
$^{2}$SUPA, School of Physics and Astronomy, University of St Andrews, North Haugh, St Andrews, Fife, KY16 9SS, United Kingdom\\
$^{3}$Institut d'Astronomie et d'Astrophysique, Universite Libre de Bruxelles, CP226, Bvd du Triomphe, B-1050, Bruxelles, Belgium}

\begin{document}

\date{Accepted \dots . Received \dots; in original form \dots}


\maketitle

\label{firstpage}

\begin{abstract}
Strong gravitational lensing by galaxies in MOdified Newtonian Dynamics (MOND) has until now been restricted to spherically symmetric models. These models were able to account for the size of the Einstein ring of observed lenses, but were unable to account for double-imaged systems with collinear images, as well as four-image lenses. Non-spherical models are generally cumbersome to compute numerically in MOND, but we present here a class of analytic non-spherical models that can be applied to fit double-imaged and quadruple-imaged systems. We use them to obtain a reasonable MOND fit to ten double-imaged systems, as well as to the quadruple-imaged system Q2237+030 which is an isolated bulge-disc lens producing an Einstein cross. However, we also find five double-imaged systems and three quadruple-imaged systems for which no reasonable MOND fit can be obtained with our models. We argue that this is mostly due to the intrinsic limitation of the analytic models, even though the presence of small amounts of additional dark mass on galaxy scales in MOND is also plausible. 
\end{abstract}


\section{Introduction}

The MOdified Newtonian Dynamics paradigm (MOND, \citealt{Mond3}) has been originally proposed as an alternative to galactic cold dark matter (CDM). This paradigm postulates a modification of Newton's laws in order to explain the rotation curves of spiral galaxies, by saying that the MONDian gravitational accelerations $g_M \ll a_0 \approx 10^{-10} {\rm m} \, {\rm s}^{-2}$ approach $(g_N a_0)^{1/2}$ where $g_N$ is the usual Newtonian gravitational acceleration. Nowadays, this paradigm is still better suited than dark matter to explain the observed conspiracy between the distribution of baryons and the gravitational field in spiral galaxies \citep[e.g.,][]{insight2,insight}. Indeed, it is baffling that such a simple prescription leads to uncannily successful predictions for galaxies ranging over five decades in mass (see, e.g., \citealt{spiral1,mondref1} for reviews), including our own Milky Way \citep{milky,escape}. 

On the other hand, recent developments in the theory of gravity have also added plausibility to the case for MOND through the advent of Lorentz-covariant theories yielding this behavior in the ultra-weak gravity limit by means of a dynamical (not necessarily) normalized four-vector field \citep{teves,tv2,fieldtheo,vectornew,covariant1,covariant2}. Although rather fine-tuned and still being a far cry from a fundamental theory underpinning the MOND paradigm, these new theories remarkably allow for new predictions regarding cosmology \citep[e.g.,][]{tevesneutrinocosmo,dod1,dod2,dod3} and gravitational lensing \citep[e.g.,][]{chiu,qin,lenstest,tevesfit,wedding,chiba,filamentlens}.

In these Lorentz-covariant theories, gravitational lensing has actually been shown to work exactly in the same manner as in General Relativity (GR) for static systems \citep[e.g.,][]{teves}, except for the relation between the density and the gravitational potential. This allowed, for instance, \cite{lenstest} to test MOND against the CfA-Arizona Space Telescope Lens Survey (CASTLES) data of image-splitting lenses, building on the earlier work of \citeauthor{qin}\citep[1995; also see][]{extralenstest}. They were nevertheless restricted to models of {\it spherical geometry}, and were thus only able to account for the size of the Einstein ring of observed lenses, but not for the exact position of collinear images in double-imaged systems, and of course not for quadruple-imaged systems. This intrinsic limitation is due to the fact that the MONDian acceleration $\vec{g}_M$ is related to the Newtonian one, $\vec{g}_N=- \vec\nabla\Phi_{N}$, by \citep{mondnew}
\begin{equation}
\mu\left(\frac{|\vec{g}_{M}|}{a_{0}}\right)\vec{g}_{M} = - \vec\nabla\Phi_{N} +\vec S,
\label{eq:A}
\end{equation}
where $\vec S$ is a solenoidal vector field determined by the condition that $\vec{g}_M$ can be expressed as the gradient of a scalar potential $\Phi_M$. The function $\mu$, controlling the modification of Newton's law, has the following asymptotic behavior:
\begin{equation}
\begin{split}
\mu(x) \sim x  \qquad x \ll 1,\\
\mu(x) \sim 1  \qquad x \gg 1.
\end{split}
\label{eq:A2}
\end{equation}
The field $\vec{S}$ is typically non-zero in non-spherical geometry, and calculating it usually requires the use of a numerical solver for the non-linear field equation of MOND \citep[e.g.,][]{numericmond,numericmond2,asymmetric}.

In this paper, we will demonstrate how to create simple analytic models of {\it non-spherical lenses} with $\vec{S}=\vec{0}$ in MOND, choosing the
covariant framework of tensor-vector-scalar gravity \citep[TeVeS,][]{teves}. Without resorting to a numerical Poisson solver, these analytic models can thus be used to fit image positions in double-imaged and quadrupled-image systems. We present our analytic models in Sect.~2, devise a fitting procedure in Sect.~3, and finally expose and discuss the results in Sects.~4 and~5.

\section{The Hernquist-Kuzmin Model}

\subsection{Potential-density pair}
The Kuzmin disc \citep{kuzmin}, defined by a Newtonian gravitational potential of the form
\begin{equation}
\Phi_{N,K} = \frac{-GM}{\sqrt{x^2+y^2+(|z|+b)^2}} \; , \; b>0
\label{eq:B}
\end{equation}
is a well-known non-spherical case which satisfies $\vec{S}=\vec{0}$ in Eq. \eqref{eq:A}: For $z>0$, Eq. \eqref{eq:B} corresponds to the Newtonian potential generated by a point mass located at $(0,0,-b)$, in case of $z<0$ it turns into the Newtonian potential of a point mass located at
$(0,0,b)$. Thus, above and below the disc, we effectively have a spherical Newtonian potential, which implies that truly $\vec{S}=\vec{0}$ in Eq. \eqref{eq:A}.  

Hereafter, the idea is simply to model lens galaxies by replacing 
the auxiliary point lens potential of the Kuzmin disc with an auxiliary Hernquist lens potential \citep{hernquist}; we shall refer to this model as the Hernquist-Kuzmin (HK) model. A similar approach, using Plummer's model and a smooth transition at $z=0$ instead, leads to the Plummer-Kuzmin model derived by \cite{miyamoto} which provides a qualitatively good fit to the mass profile of observed galaxies. Although our proposed model is not a very good description of real galaxies, it enables us to derive fully analytic lens models in the context of MOND (see Sect. \ref{lensing}) and to study the influence of non-sphericity on the ability to fit image positions.

The Newtonian potential of the Hernquist-Kuzmin model takes the form
\begin{equation}
\Phi_{N,HK}=\frac{-GM}{\sqrt{x^2+y^2+(|z|+b)^2}+h},
\label{eq:C}
\end{equation}
with $b$ being the Kuzmin parameter and $h$ denoting the core radius of the Hernquist profile. Choosing 
different ratios $h/b$, this model will produce different Hubble type galaxies, going from a pure Kuzmin disc
galaxy for $h/b \rightarrow 0$ to a pure Hernquist sphere for $h/b \rightarrow \infty$. To clarify this situation and to characterize the non-sphericity of the model, one may simply expand the r.h.s of Eq. \eqref{eq:C} far away from the disc ($r^2=x^2+y^2+z^2$):
\begin{equation}
\Phi_{N,HK}=\frac{-GM}{r+h}\left( 1 - {\frac {\left| z \right| b}{(r+h)r}}\right)+\mathcal{O} \left( {b}^{2}
 \right)
\label{eq:C1}
\end{equation}

\begin{figure*}
 \centering
   \begin{minipage}[t]{8cm}
\begin{center} 
\includegraphics[trim=0 0 0 0,width=0.65\textwidth]{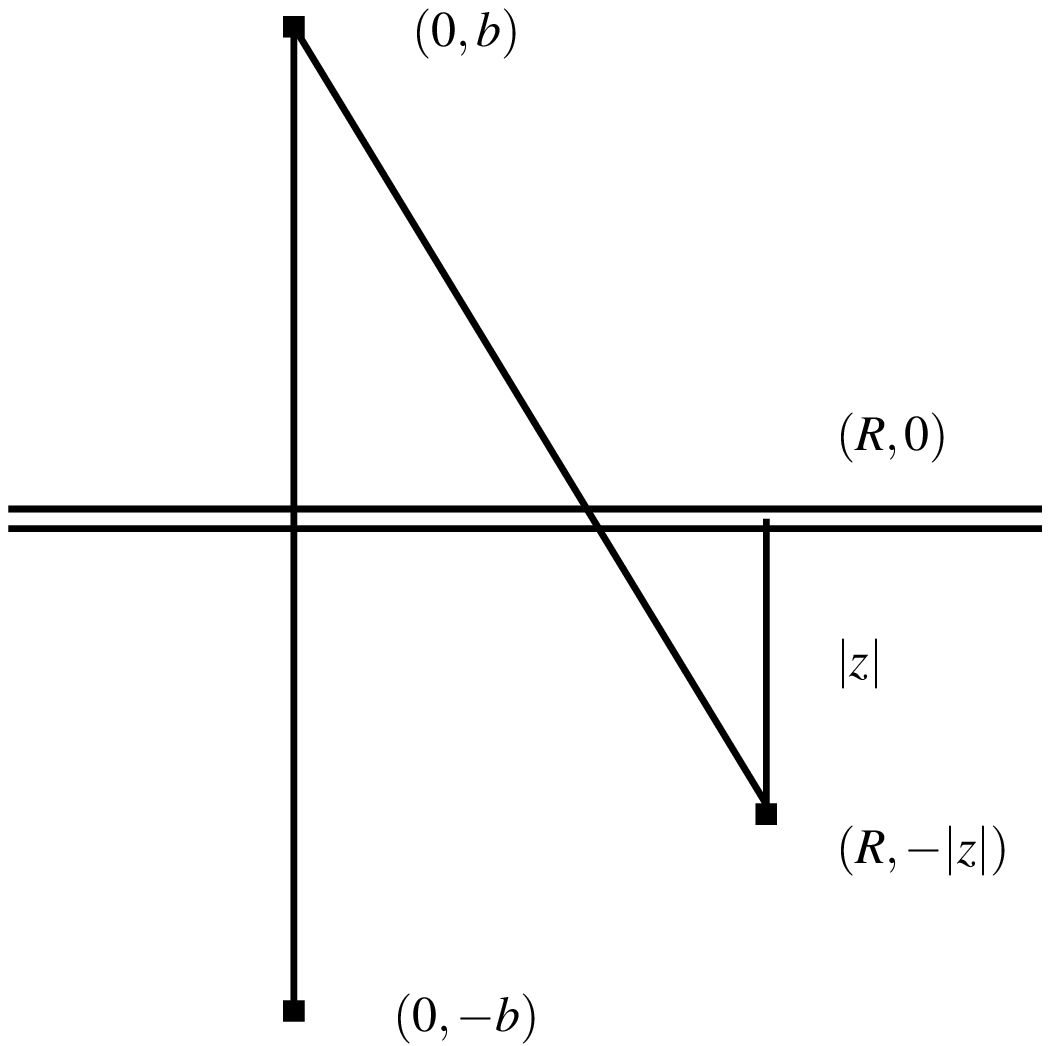}\\[0.4cm]
\includegraphics[trim=0 0 0 0,width=0.75\textwidth]{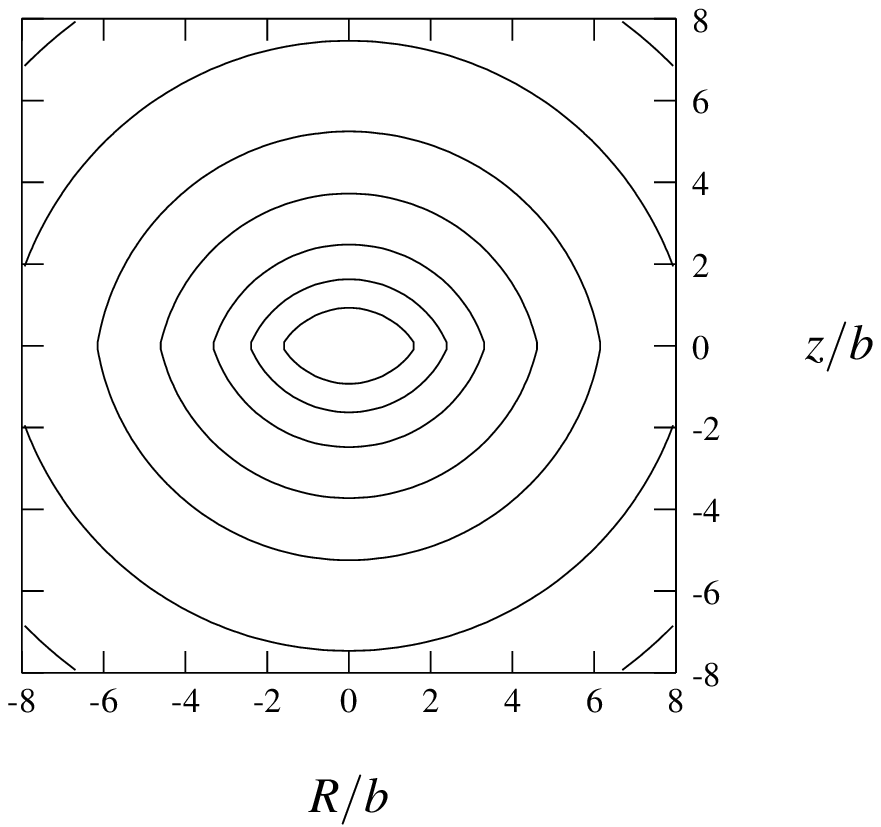}
\end{center}
  \end{minipage}
\qquad
 \begin{minipage}[t]{8cm}
\begin{center}
\includegraphics[trim=0 0 0 0,width=0.75\textwidth]{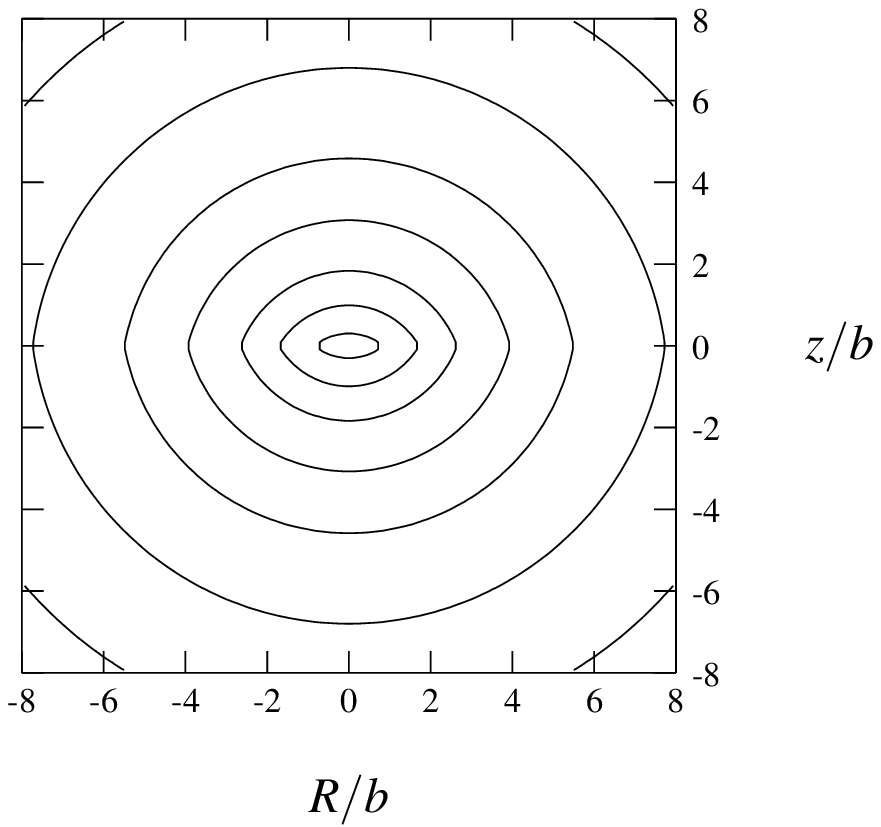}\\[0.4cm]
\includegraphics[trim=0 0 0 0,width=0.75\textwidth]{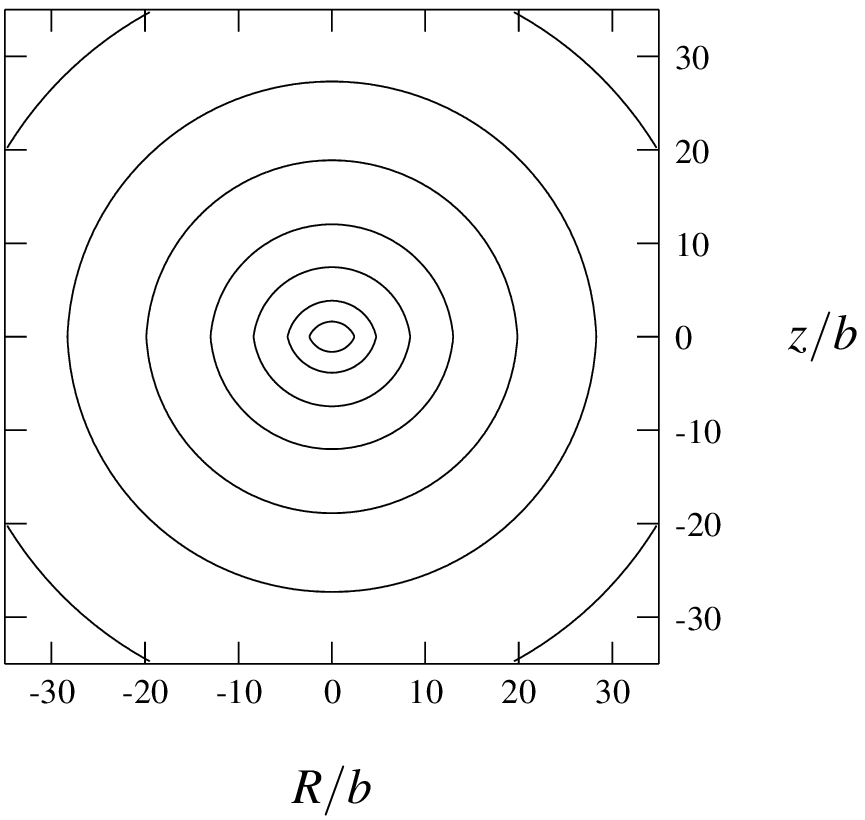}
\end{center}
\end{minipage}
\caption{Contours of equal density in the $(R,z)$ plane for the HK model \eqref{hkdensity} when $h/b=0.1$ (top right), $h/b=1$ (bottom left) and $h/b=10$ (bottom right). Contour levels are $(0.01, 0.003, 0.001, ...)M/b^3$ (top right); $(0.001, ...)M/b^3$ (bottom left); $(0.0003, ...)M/b^3$ (bottom right). The top left panel illustrates the HK model: At the point $(R,-|z|)$ below the disc, the potential Eq. \eqref{eq:C} is identical with that of a Hernquist distribution whose origin is located at a distance $b$ above the disc's centre.}
\protect\label{contour}
\end{figure*}

Using Poisson's equation, we find that the underlying density distribution is given by ($R^2=x^2+y^2$)
\begin{equation}
\rho_{HK}=\dfrac{Mh}{2 \pi \sqrt{R^2+(|z|+b)^2} \left (\sqrt{R^2+(|z|+b)^2}+h\right )^3}.
\label{hkdensity}
\end{equation}
The corresponding density contours in the $(R,z)$ plane are plotted in Fig.~\ref{contour} for different values of $h/b$.

Considering the Hernquist-Kuzmin model for gravitational lensing, we choose the $z$-axis such that it is parallel to the line-of-sight and $x$,$y$ are the Cartesian coordinates spanning the lens plane. Because we need to account for different possible orientations of galaxies, we additionally have to rotate the disc. Defining 
$(r^{'})^2=(x^{'})^2+(y^{'})^2+(|z^{'}|+b)^2$, where
\begin{equation}
\begin{split}
x^{'} &= (x\cos\phi-y\sin\phi)\cos\theta-z\sin\theta,\\
y^{'} &= x\sin\phi + y\cos\phi,\\
z^{'} &= (x\cos\phi-y\sin\phi)\sin\theta+z\cos\theta,
\end{split}
\label{eq:C2}
\end{equation}
the angle $(\pi/2)-\theta$ being the inclination of the galaxy's symmetry plane w.r.t. the line of sight and $\phi$ the galaxy's position angle (PA), Eq. \eqref{eq:C} turns into
\begin{equation}
\Phi_{N,HK} = \frac{-GM}{r^{'}+h}.
\label{eq:C3}
\end{equation}

\subsection{Lensing Properties}
\label{lensing}
In this work, we consider MOND within the framework of TeVeS \citep{teves}, henceforth using units with $c=1$.
Since most of the light bending still occurs within a small range around the lens compared to the distances between lens and source
and observer and source, gravitational lensing works exactly the same way as in GR if the Newtonian potential is substituted by the total non-relativistic modified potential.

Then, according to \cite{gl}, the resulting deflection angle can be written as
\begin{equation}
\vec{{\alpha}} = 2\int\limits_{-\infty}^{\infty}\vec\nabla_{\bot}\Phi dz,
\label{eq:E}
\end{equation}
where $\Phi$ is the total gravitational potential, $\vec\nabla_{\bot}$ denotes the two-dimensional gradient operator perpendicular to light propagation and integration is performed along the unperturbed light path (Born's approximation). Positions in the source and the lens plane are related through the lens equation:
\begin{equation}
\vec\eta = \frac{D_s}{D_l}\vec\xi - D_{ls}\vec{{\alpha}}(\vec\xi).
\label{eq:F}
\end{equation}
Here $\vec\xi=(x,y)$, $\vec\eta$ is the 2-dimensional position vector in the source plane, and $D_{s}$, $D_{l}$, and $D_{ls}$ are the (angular diameter) distances between source and observer, lens and observer, and lens and source, respectively.
Furthermore, it is convenient to introduce the deflection potential $\Psi(\vec\theta)$:
\begin{equation}
\Psi(\vec\theta) = 2D\int\Phi(D_{l}\vec\theta,z)dz,
\label{eq:0b}
\end{equation}
where we have used $\vec\theta=\vec\xi/D_{l}$ and $D=D_{ls}/(D_{s}D_{l})$. If a source is much smaller than the angular scale on which the lens properties change, the lens mapping can locally be linearized. Thus, the distortion of an image can be described by the Jacobian matrix
\begin{equation}
\mathcal{A}(\vec\theta) = \frac{\partial\vec\beta}{\partial\vec\theta} =
\begin{pmatrix}
1-\kappa-\gamma_{1} & -\gamma_{2}\\
-\gamma_{2} & 1-\kappa+\gamma_{1}
\end{pmatrix},
\label{eq:0c}
\end{equation}
with the angular coordinate $\vec\beta$=$\vec\eta/D_{s}$. The convergence $\kappa$ is directly given by the deflection potential $\Psi$:
\begin{equation}
\kappa = \frac{1}{2}\Delta_{\vec\theta}\Psi .
\label{eq:0d}
\end{equation}
Similarly, the shear components $\gamma_{1},\gamma_{2}$ can be calculated from
\begin{equation}
\begin{split}
\gamma_{1} &= \frac{1}{2}\left(\frac{\partial^{2}\Psi}{\partial\theta_{1}^{2}}-\frac{\partial^{2}\Psi}{\partial\theta_{2}^{2}}\right),\quad\gamma_{2} = \frac{\partial^{2}\Psi}{\partial\theta_{1}\partial\theta_{2}},\\
\gamma &= \sqrt{\gamma_{1}^{2}+\gamma_{2}^{2}}.
\end{split}
\label{eq:0e}
\end{equation}
Due to Liouville's theorem, gravitational lensing preserves the surface brightness, but it changes the apparent solid angle of a source. The resulting flux ratio between image and source can be expressed in terms of the amplification $A$,
\begin{equation}
A^{-1} = (1-\kappa)^{2}-\gamma^{2}.
\label{eq:0f}
\end{equation}
Consequently, the flux ratio between two images A and B is $f_{AB}=A_A/A_B$. Points in the lens plane where $A^{-1}=0$ form closed curves, the critical curves. Their corresponding curves located in the source plane are called caustics. Images near critical curves can significantly be magnified and distorted, which, for instance, is indicated by the giant luminous arcs formed from source galaxies near caustics.

Due to the deflection by the gravitational potential, light rays traveling from a source to an observer at redshift $z=0$ will be delayed in time:
\begin{equation}
t(\vec\theta)=\frac{1+z_{l}}{D}\left( \frac{1}{2} (\vec\theta-\vec\beta )^2-\Psi(\vec\theta)\right ),
\label{timedelay}
\end{equation}
where $z_{l}$ is the lens' redshift. If the deflection potential is known, Eq. \eqref{timedelay} allows to calculate the relative time delay
between different images.

Choosing a certain smooth transition from Newtonian dynamics to MOND \citep[see][Sect. 8.2 for details]{lenstest} and assuming spherical symmetry, we have the following relation for the total acceleration in TeVeS:
\begin{equation}
g_M(r)=g_N(r)+\sqrt{g_N(r)a_0},
\label{eq:Z}
\end{equation}
where $a_0=1.2 \times 10^{-10} {\rm m} {\rm s}^{-2}$. Exploiting the above and introducing $z_{0}=(x\cos\phi-y\sin\phi)\tan\theta$, the deflection angle's $x$-component yields
\begin{equation}
\begin{split}
\alpha_{x} &= 2(x-b\cos\phi\cos\theta )\int\limits_{-\infty}^{z_{0}}\frac{dz}{r^{'}}\left(\frac{GM}{(r^{'}+h)^2}+\frac{\sqrt{GMa_{0}}}{r^{'}+h}\right)\\
&+2(x+b\cos\phi\cos\theta )\int\limits_{z_{0}}^{\infty}\frac{dz}{r^{'}}\left(\frac{GM}{(r^{'}+h)^2}+\frac{\sqrt{GMa_{0}}}{r^{'}+h}\right).
\end{split}
\label{eq:Y}
\end{equation}
The integral \eqref{eq:Y} can be evaluated by means of elementary calculus, but as the resulting expression is quite lengthy, we shall skip its presentation at this point. Analogously, the closed analytic form for $\alpha_{y}$ can be derived, and as as a consequence, this is also true for the
lensing quantities $\kappa$ and $\gamma$.

\subsection{Cosmology and distances}

Concerning the calculation of distances in gravitational lensing, we shall adopt a standard flat $\Lambda$CDM cosmology with $\Omega_{m}=0.3$ and
$\Omega_{\Lambda}=0.7$. This choice is justified by the fact that many covariant formulations of MOND mimic the behavior of a $\Lambda$CDM model,
accounting for marginal differences that will have no significant impact on our analysis.
For instance,  \cite{lenstest} presented a minimal-matter open model with a cosmological constant which works to good accuracy for redshifts $z\lesssim 3$.  
Furthermore, \cite{tevesfit} proposed the flat $\mu$HDM (MOND Hot Dark Matter) cosmological model based on the assumption of massive (sterile) neutrinos.   If provided with a cosmological constant, this model is quite similar to the standard flat $\Lambda$CDM model, being capable of explaining the cosmic microwave background \citep{tevesneutrinocosmo}. Recent work has also shown that it seems possible to construct certain covariant formulations of MOND giving a cosmology identical to $\Lambda$CDM in the matter-dominated era \citep{vectornew}.

\section{Fitting procedure}
We will follow \cite{chi1} to model the lens systems: For each pair of images $i$ and $j$,  when tracing one light-ray back for each observed image to the source plane, the source position obtained from Eq. (\ref{eq:F}) should be the same for both images. We can thus simply compare the resulting source position for each image by computing their squared deviation,
\begin{equation}
\Delta^2_{s}=\sum_{i \ne j}\left( (x_{si}-x_{sj})^2 + (y_{si}-y_{sj})^2 \right),
\label{eq:J}
\end{equation}
where $x_{s}$ and $y_{s}$ is the source position in Eq. (\ref{eq:F}).  This is a measurement of how well the images retrace back to a single point in the source plane.

Another quantity to minimize is the deviation of the lens centre from the observed optical centre, given by
\begin{equation}
\Delta^2_{l}=\left( (x_{l} )^2 + (y_{l})^2 \right),
\label{eq:J1}
\end{equation}

However, our model has generally nine fitting parameters (the lens mass $M$, the Kuzmin length $b$, the Hernquist length $h$, the PA angle, the inclination $i$, the source position $x_s, y_s$, and the lens position $x_l, y_l$), while for a double-imaged system we have only four constraints from the two image positions, and another two constraints from the observed lens optical centre. The problem is thus ill-posed. 

To cure this, and ensure the uniqueness of the solution, we use a regularization method by adding a regularization term in the minimization. This term is penalizing solutions deviating from the Fundamental Plane as well as face-on\footnote{As there is strong observational evidence supporting that the system B0218+357 corresponds to a nearly face-on spiral galaxy \citep[e.g.][]{gundahlhjorth,spiralquestion}, we choose the regularization term for this particular lens such that edge-on solutions are penalized instead. Further relaxing the penalties w.r.t. the Fundamental Plane and the observed flux ratio in Eq. \eqref{eq:M}, the fit substantially improves, corresponding to a factor of $20$ in $\Delta_{s}$.} and discy solutions, and solutions with an anomalous mass-to-light ratio or a large flux anomaly:
\begin{equation}
\begin{split}
\mathcal{P}&= \left[ (\log FP)^2 + (\cos i)^2 + \left(\dfrac{b}{b+h}\right)^2 \right] \\
& + \left[\log \dfrac{f_{AB}}{f^{\rm obs}} \right]^2 +\left[\log \dfrac{M}{M_{\rm stellar}} \right]^2.
\end{split}
\label{eq:M} 
\end{equation} 
The deviation from the Fundamental Plane 
is measured by $\log FP = \log (h/h_1) - 1.26 \log(M/M_1)$, 
and $h_1=0.72{\rm kpc}$ and $M_1=1.5\times 10^{11} \msun$ \citep{stronglensprob}.

We choose a very small regularization parameter $\lambda \sim (0.003\arcsec)^2$, and minimize the following regularized ``$\chi^2$-like" quantity,
\begin{equation}
\eta^2 = \Delta^2_{s} +\Delta_{l}^2 + \lambda \mathcal{P},
\label{eq:M2}
\end{equation}
for 14 double-imaged systems and four quadruple-imaged systems in CASTLES. Note that we also check that our results are insensitive to the detailed choice of the regularization parameter\footnote{In case of RXJ0921+4529, however, our choice of $\lambda$ creates an over-regularization effect, which results in a best-fit lens mass that is roughly by a factor $10$ smaller than estimated by \cite{lenstest} fitting the system's Einstein ring size. Decreasing the regularization parameter to $\lambda \sim (3\times 10^{-4}\arcsec)^2$ is able to resolve this issue, with $\Delta_{s}$ dropping by a factor $10$ and the lens mass now being in accordance with the previous estimate of \cite{lenstest} (see Table \ref{table1}).} and that, due to the sufficient amount of constraints (position of lens and images), the fitting procedure for quadruple-image lenses is performed with $\lambda=0$. The results are shown in Table \ref{table1} and Table \ref{table2}, respectively. Finally, note that the observed mass of each lens was calculated according to Sect. 7.1 of \citeauthor{lenstest} \citeyearpar[Eqs. (73)-(75)]{lenstest}.

\begin{table*}
\centering
\caption{\rm Fitting results for selected two-image lens systems from the CASTLES sample: In 
the table, the 
observed lens mass $M_{*}$ is calculated { according to Sect. 7.1 of \citeauthor{lenstest} \citeyearpar[Eqs. (73)-(75)]{lenstest}}, the parameter $r_h$ is the Hernquist length expected from half-light measurements \citep[values are taken from ][]{lenstest}.
We do not give $\eta^2$, but instead we list $\Delta_s$ and compare the inferred values of PA, the Hernquist length $h$, mass and flux ratio
to observations. Additionally, we predict inclination and time delay for the particular lens models.
Outliers are characterized by large differences between predicted and observed flux ratios and/or anomalous mass ratios $M/M_{*}$ (deviation larger than a factor of $3$) like, for instance, in case of RXJ0921+4529 which resides in a cluster. Note that the fitted lens position is given by $(x_{l},y_{l})\approx (0,0)$ for all lenses.}
\begin{minipage}{170mm}
\begin{center}
\begin{tabular}{@{}lccccccccc@{}}
\hline
\hline
Lens &$z_l$&  $b/h$ & $h$/$r_h$ & ${\rm M(fit/obs})$ & PA & incli. & $\Delta_s$ & $f_{AB}$(fit/obs) & $\delta t(fit/obs))$\\
\hline
   &     &   & kpc &$10^{11}M_{\odot}$   & degree & degree & ${\rm arcsec}$ &   &  days \\
\hline
Q0142-100 & 0.49& 0.25 &1.34/1.6 & 1.70/4.08 & 72.2 & 90.0 &$2.43 \times 10^{-4}$ & 8.06/8.22 &  151.5/- \\
B0218+357 &0.68& 1.0 &2.19/1.8 & 2.69/2.67 & -22.6 & 6.94 &$7.75 \times 10^{-5}$& 0.759/0.587 & 7.52/10.5\\
HE0512-3329 &0.93& 0.24 &1.45/1.8& 1.49/2.91\footnote{Note that \cite{lenstest} used a different value for $M_{*}$ based on a wrong magnitude in an older version of the CASTLES data set.} & 28.2 & 90.0 &$3.90 \times 10^{-6}$ & 0.0013/1.175 &  19.6/-\\
SDSS0903+5028 &0.39& 0.76 &1.83/1.8& 2.77/3.80 & -30.4 & 90.0 &$9.90 \times 10^{-4}$ & 2.29/2.17 & 135.2/-\\
RXJ0921+4529 &0.31 & 0.037 &7.59/1.8& 20.0/0.34 & 60.2 & 90.0 &$5.85 \times 10^{-4}$ & 3.623/3.591 & 167.2/-\\
FBQ0951+2635 &0.24& 0.13 &1.20/0.32& 0.47/0.31 & 60.3 & 90.0 &$1.23 \times 10^{-4}$ & 2.74/3.53 &  13.2/-\\
BRI0952-0115 &0.41& 0.055 &2.20/0.29& 0.58/0.27 & 124.1 & 90.0 &$5.04 \times 10^{-4}$ & 3.52/3.52 & 8.11/-\\
Q0957+561 &0.36& 1.55 &1.21/5.23& 6.94/8.44 & 40.0 & 90.0 &$1.97 \times 10^{-3}$ & 14.3/1.08 & 752.4/417.0\\
Q1017-207 &0.78& 0.0092 &2.39/1.19& 0.83/0.74 & 88.8 & 89.9 &$2.16 \times 10^{-4}$ & 0.73/0.72 &  29.0/-\\
B1030+071 &0.60& 0.10 &0.84/1.50& 1.85/1.66 & 29.3 & 90.0 &$8.04 \times 10^{-5}$ & 36.6/36.6 &  346.8/-\\
HE1104-1805 &0.73& 0.33 &0.58/2.48& 4.91/3.32 & 61.9 & 90.0 &$1.96 \times 10^{-3}$ & 0.35/3.85 & 321.2/-\\
B1600+434 &0.41& 0.18 &1.64/1.8& 1.01/0.40 & 36.8 & 90.0 &$2.09 \times 10^{-4}$ & 0.83/0.84 &  32.2/51.0\\
PKS1830-211 &0.89& 0.48 &2.75/1.8& 1.33/1.48 & 62.3 & 90.0 &$4.14 \times 10^{-4}$ & 157.3/157.3 & 32.7/26.0\\
HE2149-2745 &0.50& 0.026 &0.94/11.4& 1.04/2.00 & -30.0 & 90.0 &$2.30 \times 10^{-4}$ & 6.53/4.19 & 90.7/103.0\\
SBS0909+523 & 0.83 & 0.19 & 3.02/1.8& 2.92/13.5$^{{\thempfootnote}}$ & 49.2& 90.0 & $1.84 \times 10^{-3}$& 1.42/1.42& 65.9/- \\ 
\hline
\hline
\end{tabular}
\end{center}
\end{minipage}
\label{table1}
\end{table*}

\section{Fitting results}

\subsection{Double-imaged systems}
\label{double}
Setting $\Delta_s < 0.01\arcsec $ as a reasonable threshold for acceptable fits of the HK lens, Table \ref{table1} shows that our model is able to describe the observed image positions of all double-imaged systems, with quite a number of these systems yielding plausible parameters within the context of MOND/TeVeS.  Additionally, the HK model seems to be able to explain the flux ratios of these binaries in almost every case.

However, there are a few outliers which we will discuss in the following. Since the model should be capable of reproducing all observational constraints and the lens mass should have a value close to the stellar mass ($M/M_{*}\simeq 1$) in MOND/TeVeS, these are characterized by very poor fitting parameters in terms of large differences between predicted and observed flux ratios and/or anomalous mass ratios $M/M_{*}$ (deviation larger than a factor of $3$).


\subsubsection{RXJ0921+4529}
The system RXJ0921+4529 contains two $z_{s}=1.66$ quasars and a $H=18.2$ spiral galaxy located in between the quasar images. According to \cite{doublelens2}, this galaxy lens is quite likely to be a member of a $z_{l}=0.32$ X-ray cluster centreed on the observed field. Clearly, RXJ0921+4529 does not correspond to an isolated system, which complicates the situation in TeVeS and { provides a possible explanation for the extremely poor fit/mass ratio ($M/M_{*}\approx 59$)}. The presence of a cluster could have caused difficulties in fitting the lens as the impact of non-linear effects may be important. In addition, note that there are still some unresolved issues in MOND/TeVeS concerning clusters \cite[e.g.][]{sanders99,sanders03,neutrinos2,tevesfit,asymmetric}. 

\subsubsection{Q0957+561}
The gravitational lens Q0957+561 is the most thoroughly studied one in literature. The system involves a radio-loud quasar at redshift
$z_{s}=1.41$ which is mapped into two images by a brightest cluster galaxy (BCG) and its parent cluster at redshift $z_{l}=0.36$ \citep[e.g.][and references therein]{doublelens3a,doublelens3b}. It is also known that the lens galaxy has a small ellipticity gradient and isophote twist which are properties the simple HK model cannot account for. Together with the fact that the lens is embedded into a cluster, this might be a reason for the huge discrepancy between observed and predicted flux ratio in the context of modified gravity.

\subsubsection{HE1104-1805}
The lens galaxy's colors are in agreement with a high-redshift early-type galaxy, and its redshift is roughly estimated as $z_{l}=0.77$
\citep[and references therein]{doublelens4}. Concerning its lensing properties, the system HE 1104-1805 is quite uncommon in a sense that
the lens is closer the bright image, rather than the faint one. As is known from lensing within the standard GR/CDM paradigm, simple models can create
such configurations only for a narrow range of parameters due to the peculiar flux ratio. Assuming simple ellipsoidal lens models, however,
these parameters imply a large misalignment between the light and the projected density. The only possibility to align the mass with the light is to
have a shear field being approximately twice as strong as estimated from the particular lens model.

Furthermore, the observed image separation is by a factor $2-3$ larger than that of a typical lens, strongly suggesting that the separation is
enhanced by the presence of a group or a cluster. So far, however, there has been no direct observational evidence for such a structure in the lens'
surrounding area. 

Analog to the above-mentioned lens systems, the unsatisfying fit and the correspondingly inferred flux ratio might be a result of both lens environment and model limitations
(see also Sect. \ref{maxshear} and \ref{altmodel}).

\subsubsection{SBS0909+523}
{
SBS0909+532 shows two images of background quasar source at $z_{s}=1.377$ separated by $1.11\arcsec$
\citep[e.g.][and references therein]{doublelens4}. Optical and infrared HST images indicate that the lensing galaxy has a large effective
radius and a correspondingly low surface brightness. Additionally, the lens galaxy's redshift is estimated as $z_{l}=0.83$ \citep{Lubin}, and
its total magnitude in the $H$-band has been measured as $H=16.75\pm{0.74}$. Although the lens galaxy's colors are poorly measured, they seem
consistent with those of an early-type galaxy at the observed redshift.

The large uncertainties are a result of the difficulty in subtracting the close pair of quasar images \citep{doublelens4}. For instance, the
uncertainty in the $I$-band magnitude, $I=18.85\pm 0.45$, allows for a deviation of the mass estimate $M_{*}$ by a factor of roughly $2.3$
on a $2\sigma$ level, where we have again used Eqs. (73)-(74) of \cite{lenstest}. Thus, we argue that the low mass ratio (listed in Table
\ref{table1}) may be entirely due to these uncertainties in observed magnitudes, with better constrained observations possibly softening the
problem in MOND/TeVeS.
}


\subsubsection{HE0512-3329}
The system HE0512-3329 was discovered as a gravitational lens candidate in the course of a snapshot survey with the Space Telescope
Imaging Spectrograph (STIS), with the images of the quasar source being separated by $0.644\arcsec$ \citep[][and references therein]{microextinction}.
Although the lensing galaxy has not been detected yet, measurements of strong metal absorption lines at redshift $z = 0.93$, identified in the
integrated spectrum,  hint towards a damped Ly$\alpha$ system intervening at this redshift.

Analyzing separate spectra of both image components, \cite{microextinction} point out that both differential extinction and microlensing 
effects significantly contribute to the spectral differences and that one cannot be analyzed without taking into
account the other. For lens modeling purposes, the observed flux ratio can therefore only be used after correcting for both effects. Thus, the large discrepancy between predicted and observed flux ratio might be a consequence of neglecting the above mentioned effects, rather than being intrinsic to MOND/TeVeS.

\begin{table*}
\centering
\caption{\rm Fitting results for selected four-image lens systems from the CASTLES sample: Note that all positions (RA and Dec.) are given in units of arcsec. The observed PA and inclination of Q2237+030 (major-axis) are PA$=77.2^o$ and $i=64.5^o$, respectively, assuming a circular face-on disc. Replacing the auxiliary Hernquist with a Jaffe profile barely changes the numbers: inclination
and PA change by about 5 degrees, the predicted mass by roughly 10 per cent.}
\begin{tabular}{lcccc}
\hline
\hline
 & PG1115+080 & Q2237+030 & B1422+231 & SDSS0924+0219\\
\hline
$z_l$ & 0.31 & 0.04 & 0.34 & 0.39 \\
$z_s$ & 1.72 & 1.69 & 3.62 & 1.52 \\
$D_l$ (kpc) & 957.2 & 163.6 & 1020.2 & 1116.6\\
$D_s$ (kpc) & 1874.2 & 1874.0 & 1637.6 & 1867.0\\
$D_{ls}$ (kpc) & 1413.2 & 1810.8 & 1341.7 & 1252.1\\
$\rm Image \ A$ & $(-0.947,-0.690) \pm 0.003$ & $(-0.075,-0.939) \pm 0.003$ & $(0.375,0.973) \pm 0.003$ & $(-0.162,0.847) \pm 0.003$ \\
$\rm Image \ B$ & $(-1.096,-0.232) \pm 0.003$ & $(0.598,0.758) \pm 0.003$ & $(0.760,0.656) \pm 0.003$ & $(-0.213,-0.944) \pm 0.003$ \\
$\rm Image \ C$ & $(0.722,-0.617) \pm 0.003 $ & $(-0.710,0.271) \pm 0.003$ & $(1.097,-0.095) \pm 0.003$ & $(0.823,0.182) \pm 0.003$ \\
$\rm Image \ D$ & $(0.381,1.344)  \pm 0.003$ & $(0.791,-0.411) \pm 0.003$ & $(-1.087,-0.047) \pm 0.003$ & $(-0.701,0.388) \pm 0.003$ \\
\hline
$\rm Source$ & $(-0.011,0.091)$ & $(0.027,-0.0051)$ & $(0.089,0.030)$ & $(-0.024,-0.047)$ \\
$\rm Lens$ & $(-0.0011,-0.0041)$ & $(0.00066,0.00096)$ & $(-0.00093,0.0065)$ & $(0.019,-0.0051)$ \\
$\rm M$ $(M_{\rm fit}/M_{*})$ $(\times 10^{11}M_{\odot})$ & $7.80/1.23 $ & $0.78/1.19 $ & $4.83/0.77 $ & $2.80/0.32 $ \\
$\rm h$ $({\rm kpc})$ & $2.25$ & $0.44$ & $8.42$ & $1.57$ \\
$\rm b/h$ & $0.56$ & $1.85$ & $0.29$ & $2.17$ \\
PA angle (degree) & $244.8$ & $246.6$ & $117.9$ & $266.4$ \\
Inclination (degree) & $44.5$ & $30.6$ & $48.6$ & $40.5$ \\
$\Delta_s$ (arcsec) & $0.0402$ & $0.0026$ & $0.0593$ & $0.0612$ \\
Flux ratio(obs) &4.03:2.53:0.65:1 &2.62:1.64:1.30:1 &31.1:34.6:18.4:1 & 12.5:5.68:4.81:1\\
Flux ratio(fit) &3.98:4.15:1.40:1 &0.81:0.66:0.68:1 &8.56:6.53:7.51:1 & 1.66:0.69:0.86:1\\
\hline\hline
\end{tabular}
\label{table2}
\end{table*}

\begin{figure}
\centering
\includegraphics[trim= 0 0 0 0,width=0.7\linewidth]{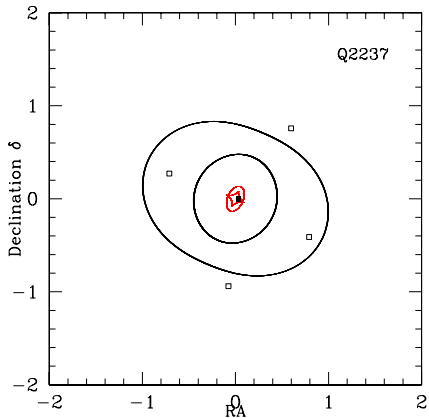}
\caption{Critical curves (black) and caustics (red) of the best-fit Hernquist-Kuzmin model for Q2237+030: The empty and filled squares denote
the observed positions of images and source, respectively.}
\label{q2237}
\end{figure}

\subsection{Quadruple-imaged systems}
\label{quad}

As we can see from Table \ref{table2}, most of the quadruple-imaged systems are very poorly fitted by our analytic HK lens model for MOND.
In accordance with our goodness-of-fit criterion ($\Delta_s < 0.01\arcsec $) introduced in Sect. \ref{double}, there is just one system where the model is able to predict the image positions in a satisfying manner. Additionally, none of the observed flux ratios can be explained.

The only acceptable fit is given for Q2237+030, the nearby Einstein cross \citep[$z_{l}=0.04$;][]{huchra}, which is the only true bulge-disc system in our set. Also, its physical Einstein ring size in the lens plane is very small, $R_{E}\approx 0.7$kpc \citep[in B1422+231, for instance, it is already by a factor of roughly $10$ larger;][]{quadeinstein}.  Nevertheless, it is not possible to give a reasonable explanation for the flux ratios using our smooth MOND lens model. Taking effects due to microlensing into account, which are not considered in this work, could be able to relax the situation. Note that the lens galaxy actually contains a bar feature \citep{barfeature} which is ignored in our analysis.

\begin{figure}
\centering
\includegraphics*[angle=270,viewport= 150 265 500 650, width=0.785\linewidth]{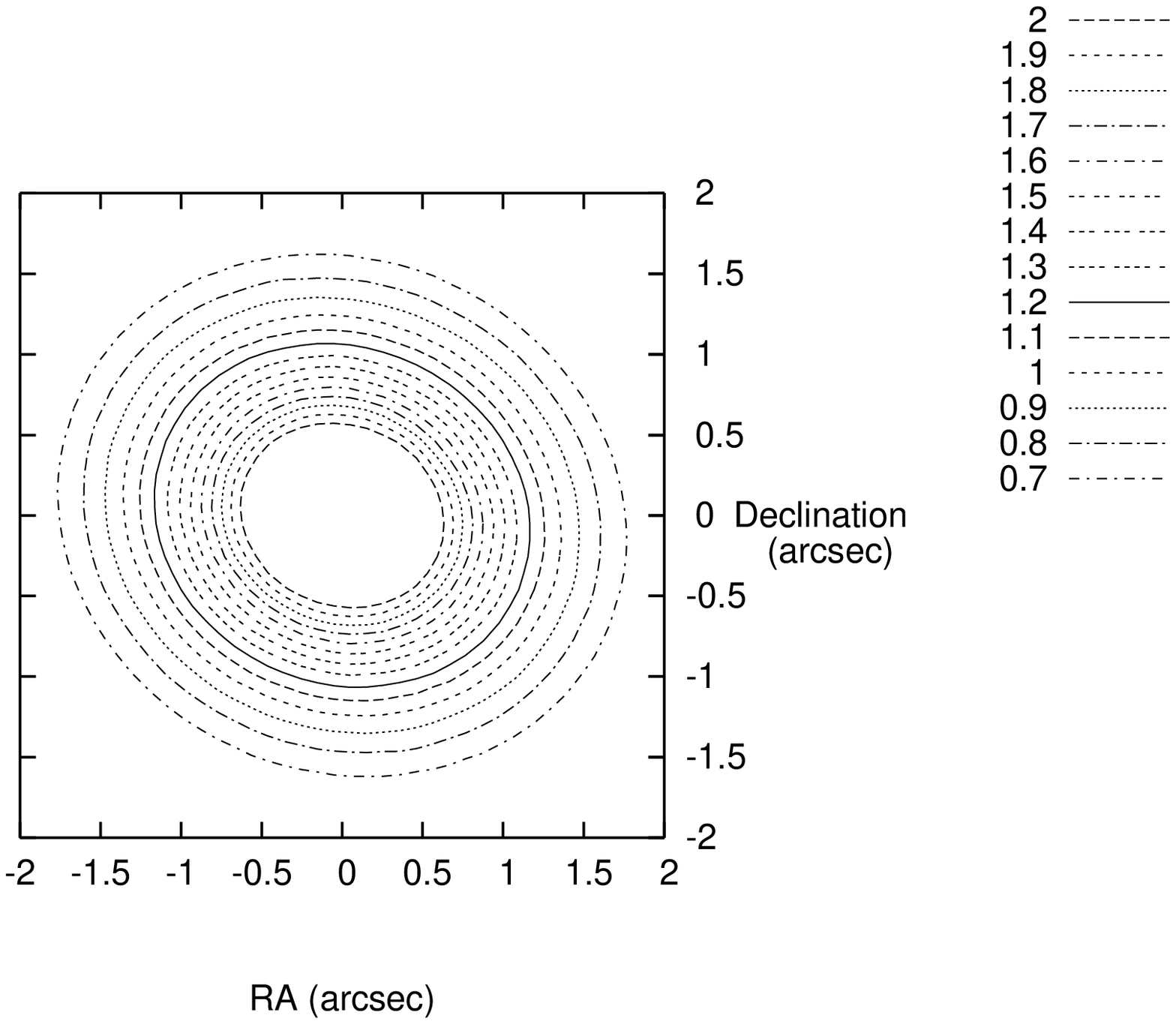}
\caption{Convergence map $\kappa$ of the best-fit Hernquist-Kuzmin model for Q2237+030, with the outer contour level being at $\kappa=0.7$ and increasing in steps of $0.1$ up to a level of $\kappa=2.0$.}
\label{q2237a}
\end{figure}

\subsubsection{PG1115+080}
The lens galaxy in PG1115+080 and its three neighbors belong to a single group at $z_{l}=0.311$, with the group being centreed southwest of
the lens galaxy's position \citep[and references therein]{quadlens1,quadlens1b}. Reasonable fits of this lens typically involve a significant
amount of external shear in the context of GR/CDM. Moreover, the observed anomaly of the flux ratio ($\sim 0.9$) between two of the images strongly
hints towards an additional perturbation of the system caused by a satellite galaxy or a globular cluster. Similar to Sect. \ref{double}, we have a
gravitationally bound system which will involve a more complicated approach in TeVeS than provided by the isolated HK model.

\subsubsection{B1422+231}
The system B1422+231 shows almost the same characteristics as PG1115+080 \citep[][and references therein]{quadlens2}. Again, the lens belongs to
a galaxy group which is centreed south of the lens galaxy ($z_{l}=3.62$). \cite{kormann1} were able to fit the lensing system using a very flat singular isothermal ellipsoid \citep[SIE;][]{minshear2,kormann2} plus an external shear field. However, HST observations revealed that the lens galaxy's optical axis ratio is much closer to unity than assumed for the flat SIE, favoring rounder lens models with larger external shear.

\subsubsection{SDSS0924+0219}
Estimated colors and magnitudes of the lens galaxy are consistent with those of a typical elliptical galaxy at $z_{l}=0.4$
\citep[and references therein]{quadlens3}.
Although the lens environment does not show any nearby objects perturbing the system, quite an amount of external shear is needed to obtain a
satisfying fit to observations, with the lens being typically modeled by a (flattened) singular isothermal sphere (SIS). Additionally, microlensing
plays an important role in explaining the observed flux ratios within GR/CDM, which is likely to be true in TeVeS as well.

\subsection{Maximum non-spherical shear of a Kuzmin lens}
\label{maxshear}
As we have seen, the outliers in our selection of quadruple-image lenses correspond to systems with a large external shear. In PG1115+080,
for example, this is due to a neighboring galaxy group.  However, the same situation also appears in uncrowded environments, 
usually constraining the lensing potential to require a substantial ellipticity. From Sect. \ref{quad}, it seems that our present analytic model is not able to generate such a potential in most cases. As is known, almost all quadruple-imaged systems show evidence for the need of an external shear field  \citep{minshear1,minshear2,minshear3} by violating a certain inequality of the image positions. It is perhaps not surprising that the current isolated HK model fails to fit these lenses\footnote{Note that our analysis does not take into account external shear effects, which would complicate the relation between lens mapping and associated density distribution due to non-linearity in modified gravity. While the main task of this paper is to explore the capability of the HK model, such contributions should certainly be addressed in future work.}.  

To gain a better understanding about this issue, we consider a pure edge-on Kuzmin lens ($h=0$) and derive the maximum variation of 
the shear at the Einstein radius $R_{E}$ by comparing its values on the major and minor axis. For this reason, let us introduce a quantity $Q$ being given as follows:
\begin{equation}
Q = \frac{\gamma(R_{E},0) - \gamma(0,R_{E})}{\gamma(R_{E},0) + \gamma(0,R_{E})}.
\label{kuzminshear}
\end{equation}
The parameter defined above will indicate the level of the shear field's non-sphericity at the Einstein radius and is a function of the dimensionless radius $R_{E}/b$. Note that, in case of the Kuzmin lens, the quantity $Q$ depends on redshift.

\begin{figure}
\includegraphics[trim= 0 0 0 0,width=0.9\linewidth]{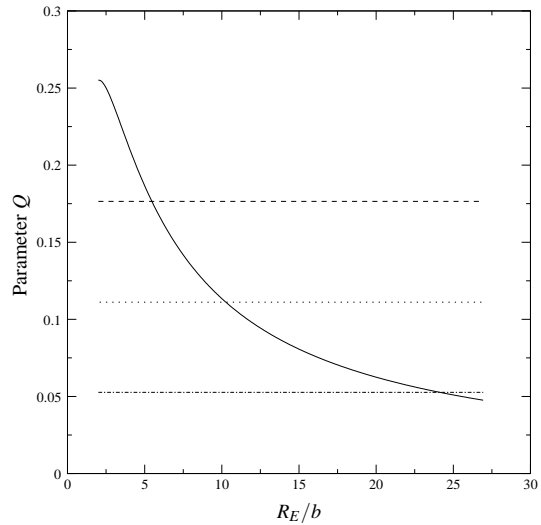}
\caption{Non-spherical shear parameter $Q$ for a simple TeVeS Kuzmin lens (solid line), assuming $a_{0}D/c^2=0.03$. Additionally, we show the results for a SIE model with a potential axis ratio of $0.9$ (dot-dashed line), $0.8$ (dotted line) and $0.7$ (dashed line), respectively.}
\label{sheartest}
\end{figure}

Figure \ref{sheartest} shows $Q$ as function of $R_{E}/b$ for the pure Kuzmin model (solid line), assuming $a_{0}D/c^2=0.03$. This value has been chosen
in accordance with the majority of lens redshifts in the CASTLES sample, and changing it does have no significant qualitative impact on the basic
outcome. Additionally, we also present the result for the SIE model \citep{kormann2}, with the potential axis ratio varying
from $0.7$ to $0.9$ (shown by horizontal lines). As we can see, the Kuzmin model becomes comparable to a very round SIE if $R_{E}/b\gtrsim 10$.

To obtain a sufficiently strong quadruple moment/non-spherical shear at the Einstein radius ($Q>0.2$), these disc-only models must satisfy
the condition $b > 0.2R_{E}$. In case of PG1115, the observed ring size can be estimated as $R_{E}\approx 5$kpc, so to fit four images, one
might actually expect that $b > r\approx 1$kpc. However, trying to fit the above mentioned Einstein ring size, using the stellar mass only,
we also find that this would need a Kuzmin parameter close to zero ($b\approx 0$), corresponding to a very concentrated point-like lens.
Although we have only given a plausibility argument, rather than a rigorous proof, this could explain why we cannot find a value of b that
meets both requirements and why the HK model mostly fails to fit quadruple-imaged systems.

\begin{figure*}
\centering
\begin{minipage}[t]{5.0cm}
\includegraphics[width=\textwidth]{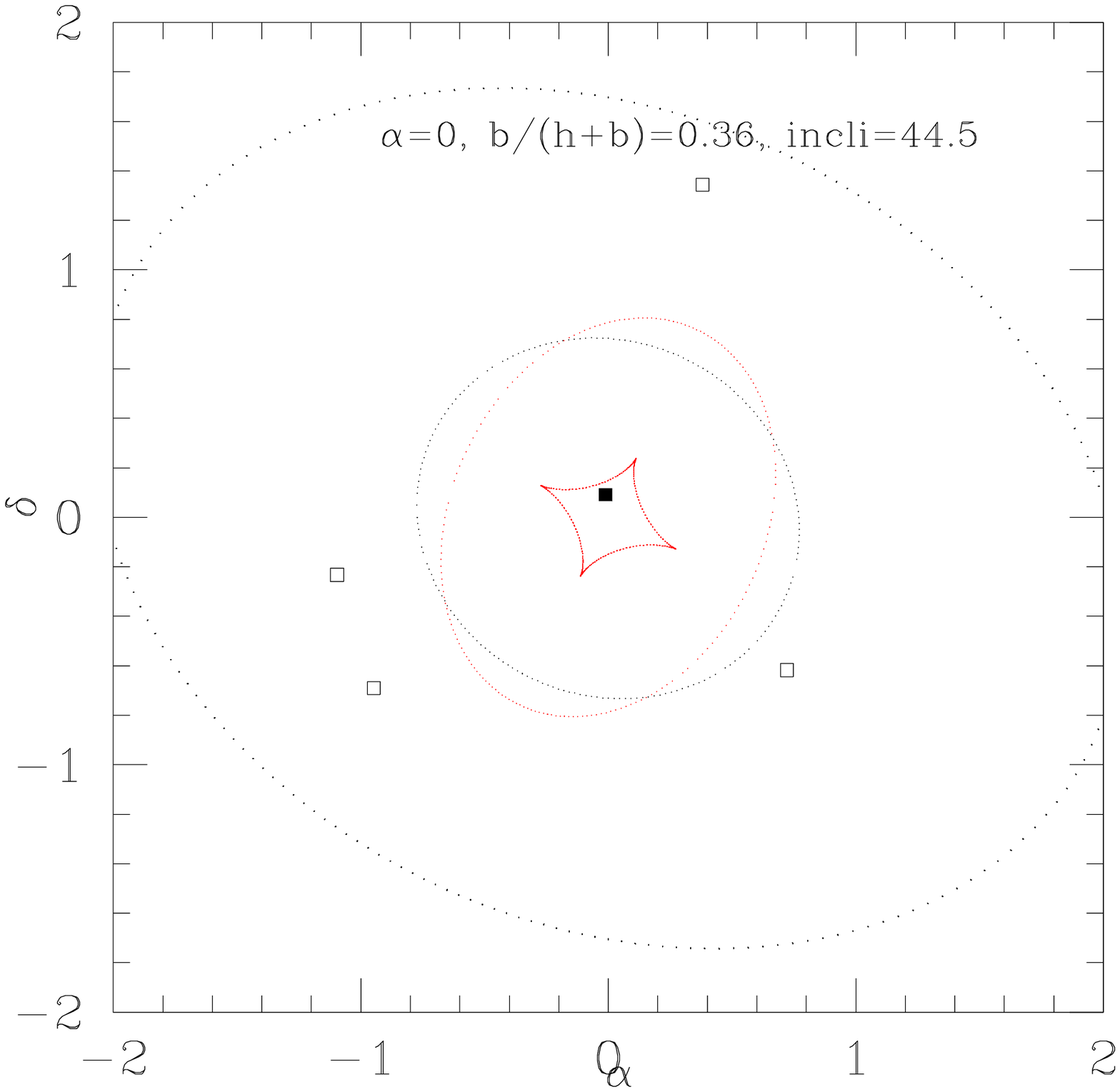}
\end{minipage}
\hfill
\begin{minipage}[t]{5.0cm}
\includegraphics[width=\textwidth]{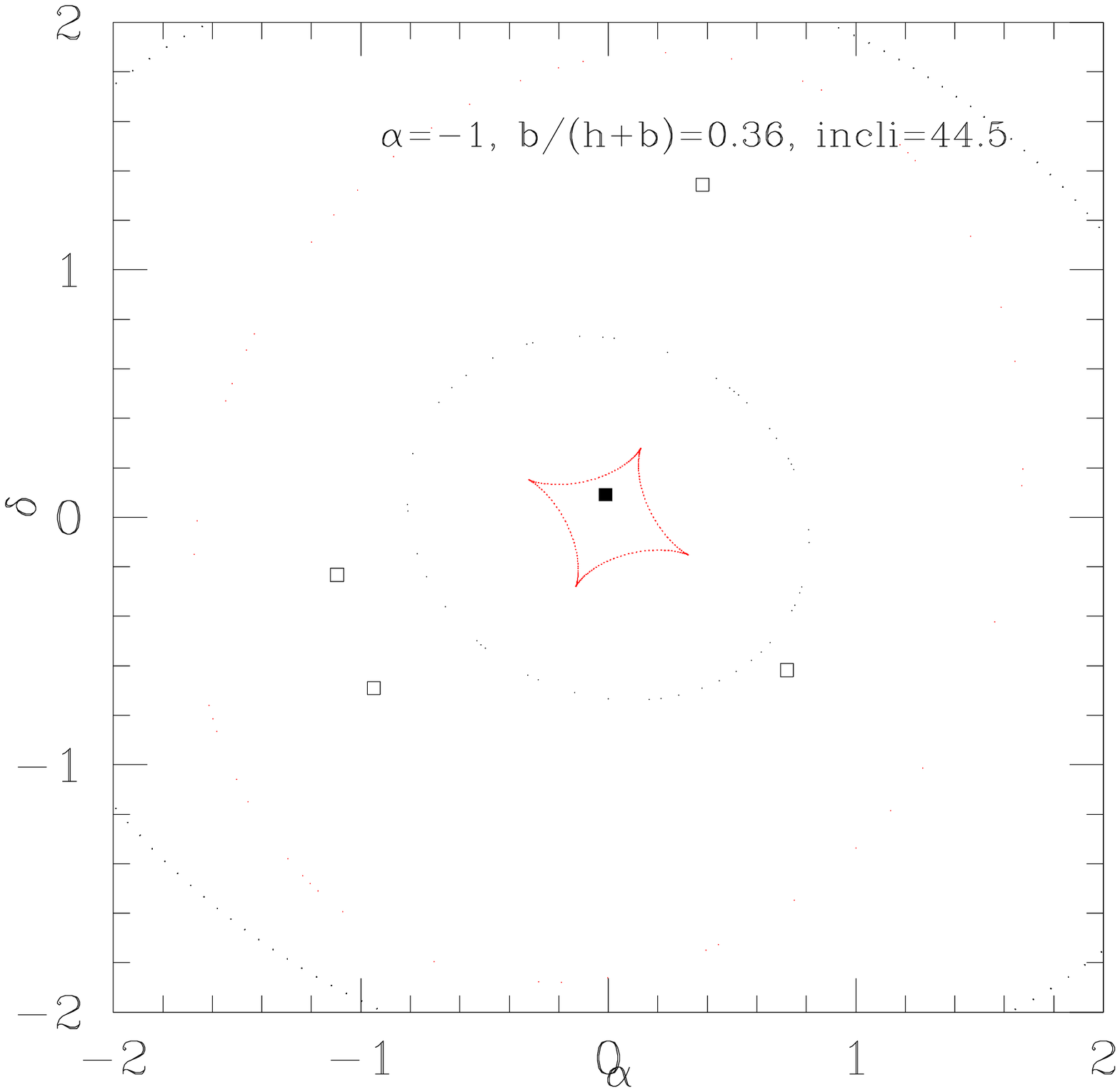}
\end{minipage}
\hfill
\begin{minipage}[t]{5.0cm}
\includegraphics[width=\textwidth]{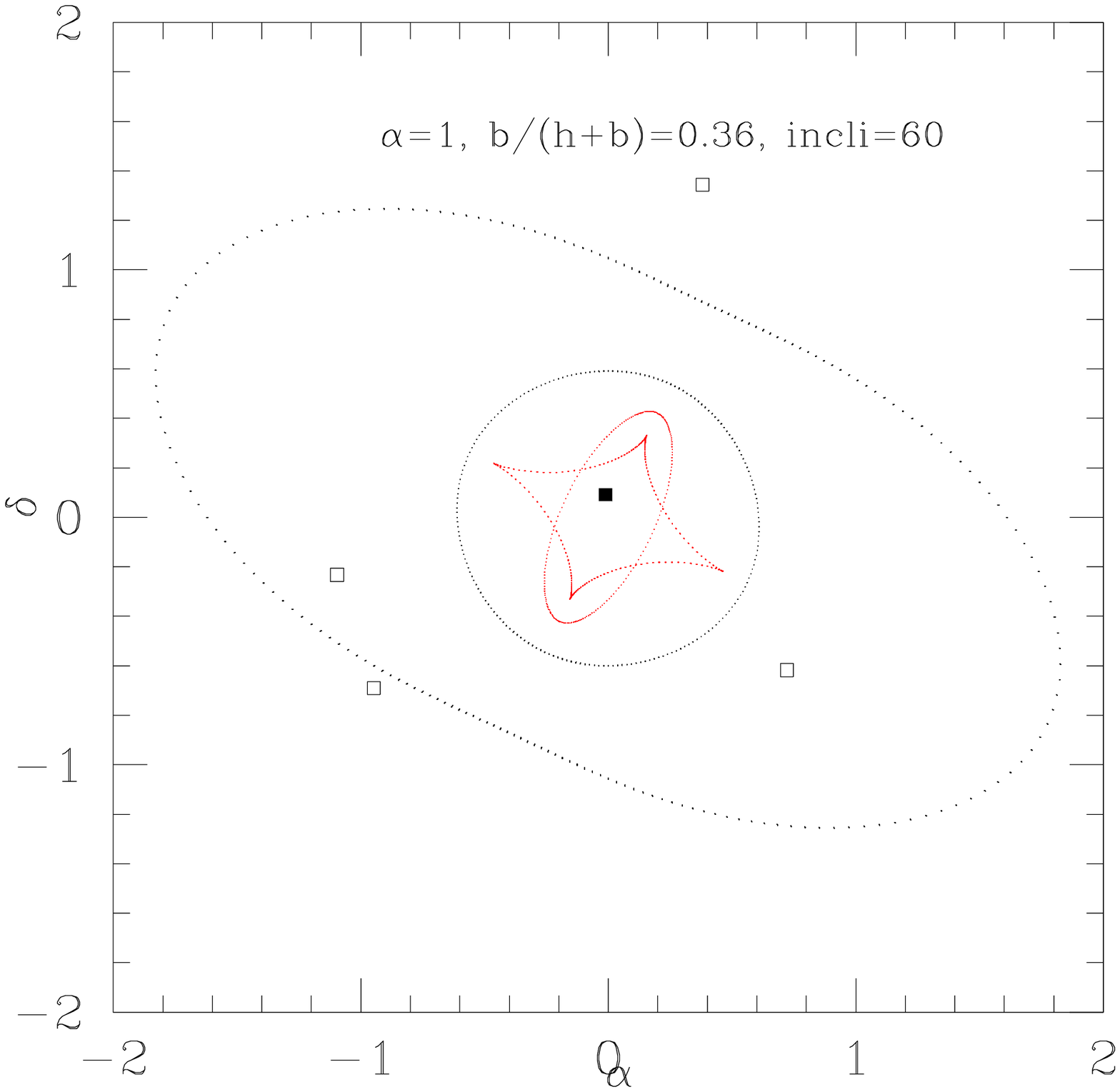}
\end{minipage}
\caption{Experimenting with hypothetical lenses: Shown are the critical curves (black) and caustics (red) of different Dehnen-Kuzmin models characterized by the parameters $\alpha$, $b/(h+b)$ and inclination (``core" radius and PA are fixed to $h=0.72$kpc and $77.2^{\circ}$, respectively). All models assume a lens mass of $M=8\times 10^{11}\Msun$ which is approximately 8 times the stellar mass of the lens galaxy in PG1115+080. The empty and filled squares denote the observed image and source positions of PG1115+080.}
\label{selection}
\end{figure*}

\begin{figure*}
\centering
\begin{minipage}[t]{5.0cm}
\includegraphics[width=\textwidth]{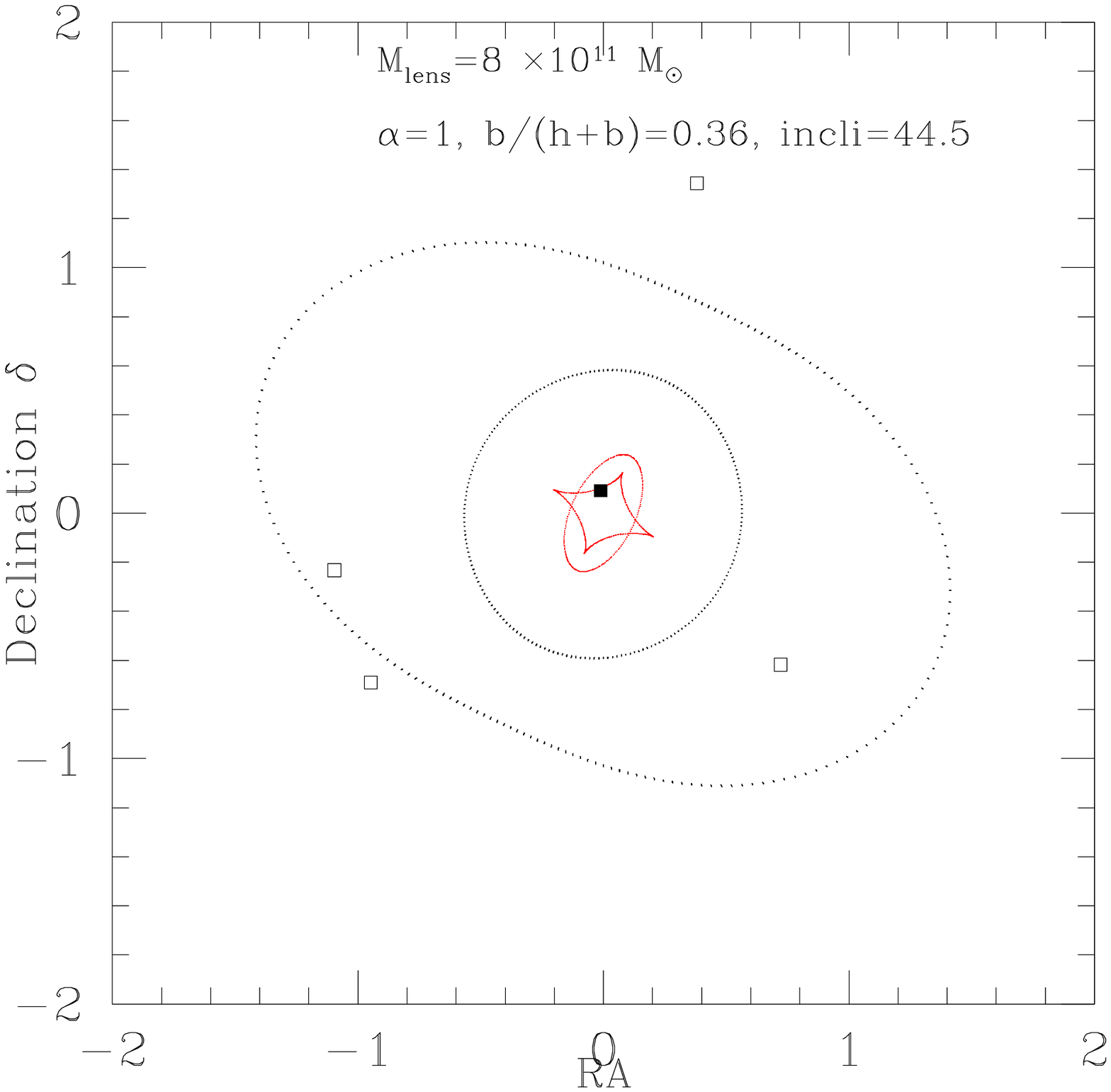}
\end{minipage}
\hfill
\begin{minipage}[t]{5.0cm}
\includegraphics[width=\textwidth]{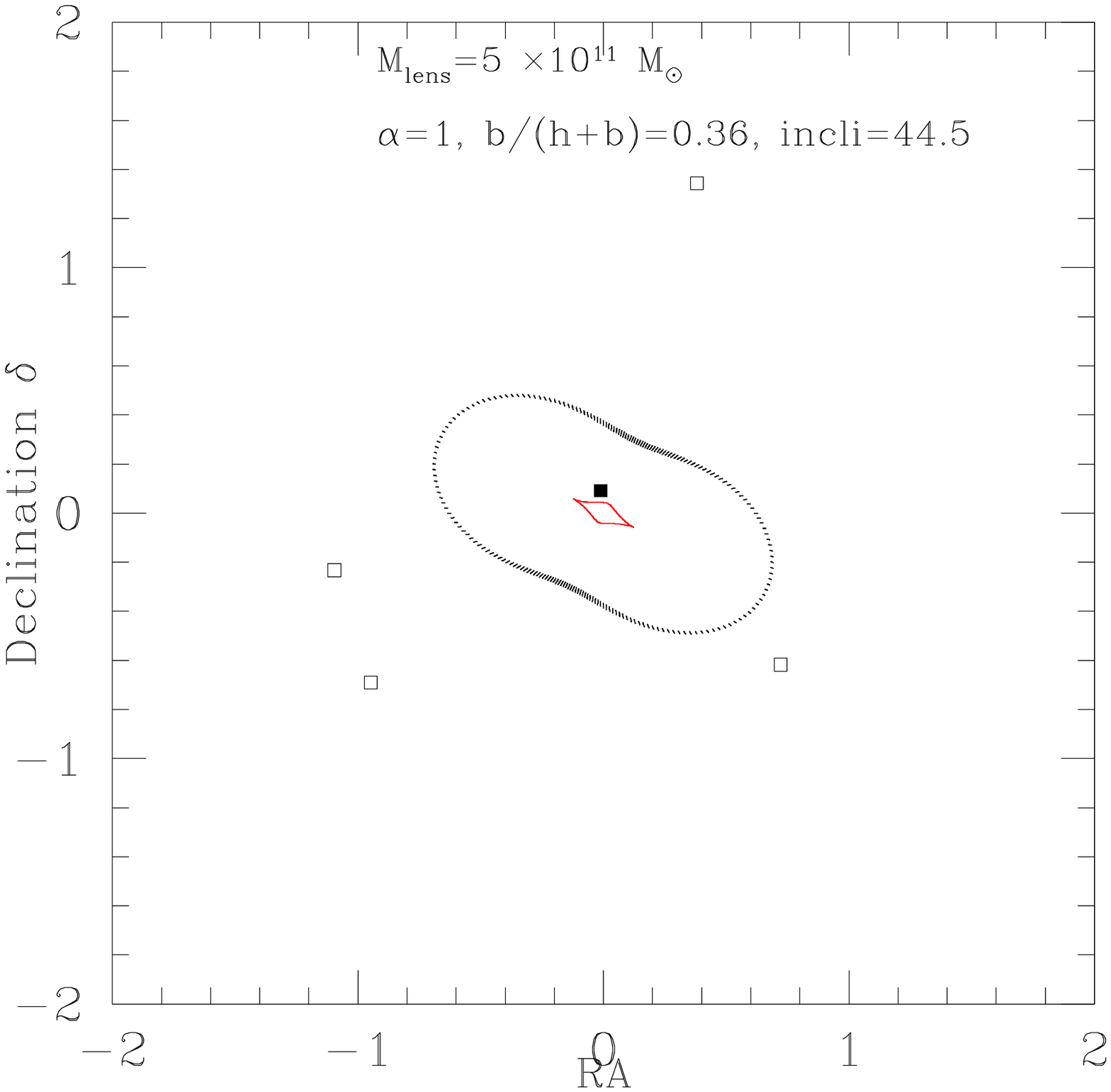}
\end{minipage}
\hfill
\begin{minipage}[t]{5.0cm}
\includegraphics[width=\textwidth]{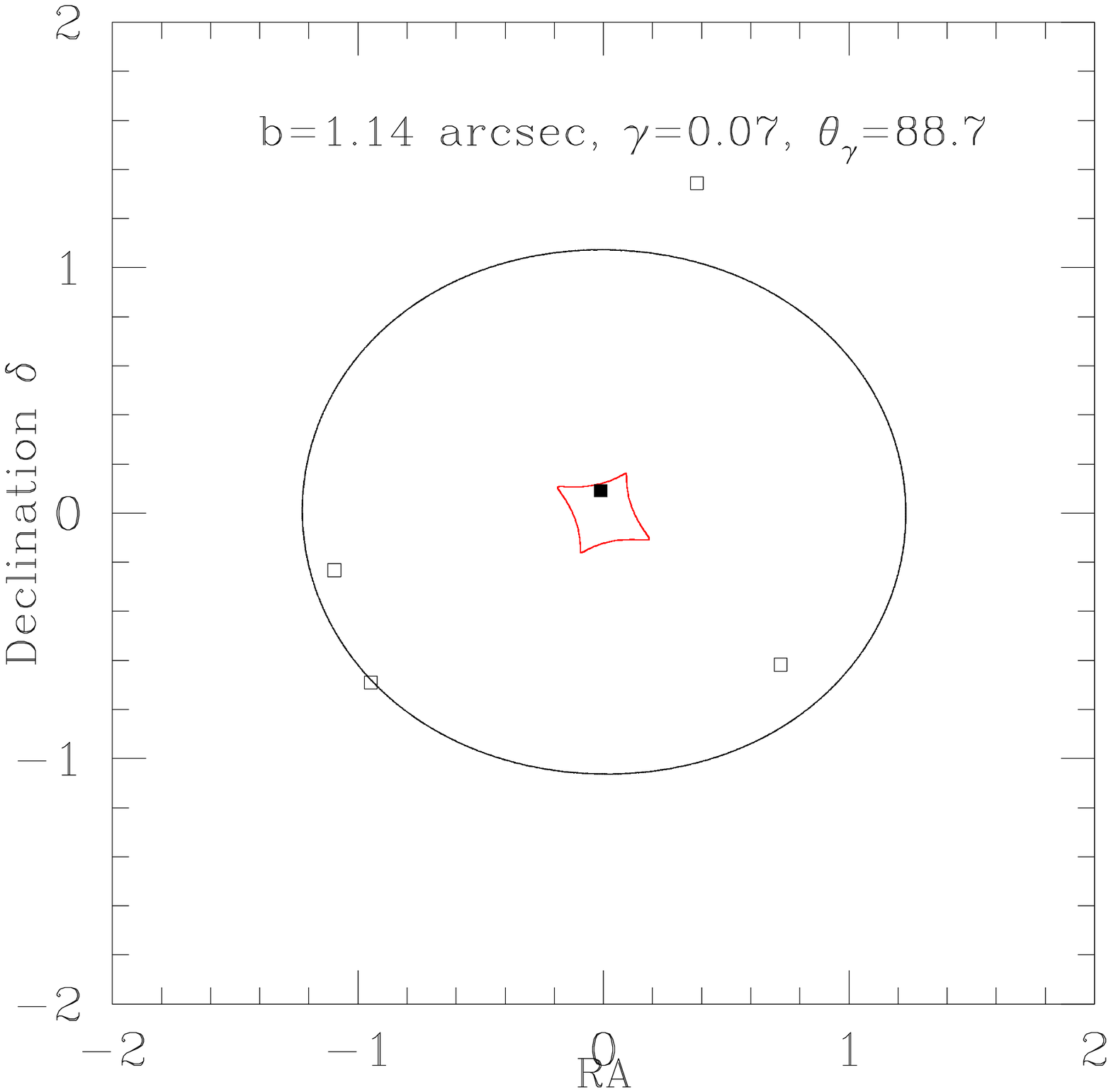}
\end{minipage}
\caption{Effects of reducing the lens mass:  Shown are the critical curves (black) and caustics (red) of a Dehnen-Kuzmin model ($\alpha=1$, $b/(h+b)=0.38$, $h=0.72$kpc, PA$=77.2^{\circ}$ and $i=44.5^{\circ}$), assuming $M=8\times 10^{11}\Msun$ (left panel) and $M=5\times 10^{11}\Msun$ (middle panel), respectively. The empty and filled squares denote the observed image and source positions of PG1115+080, the stellar mass of the lens is estimated as $M_{*}\approx 10^{11}\msun$.
Right panel: Critical curves (black) and caustics (red) of the best-fit SIS+$\gamma_{ext}$ model given by Eq. \eqref{sisext}.}
\label{SIS}
\end{figure*}

\subsection{Experimenting with hypothetical lenses}
\label{altmodel}
Another possibility of investigating the fitting capability of our model is to generally explore its parameter space and to
study the structure of critical curves and caustics. To avoid any limitations that might be due to the particularly chosen radial profile,
we furthermore replace the auxiliary Hernquist profile with the more general model proposed by \cite{dehnen}. Its Newtonian potential and
the corresponding density profile read as
\begin{equation}
\begin{split}
\Phi_{N,D}(r) &= \frac{GM}{\alpha} \left[ -1 + \left( \frac{r}{r+h} \right)^\alpha \right ],\\
\rho_{D}(r) &= \frac{M h (1+\alpha)}{ 4 \pi r^{2-\alpha} (r+h)^{2+\alpha} },
\end{split}
\label{eq:D}
\end{equation}
where $h$ is a characteristic length of the model. Depending on the value of $\alpha$, the Dehnen model represents different density distributions,
ranging from quite cuspy to more broadened profiles.  For $\alpha=0$ and $\alpha=1$, Eq. \eqref{eq:D} reduces to the models of Jaffe \citep{jaffe} and Hernquist, respectively.

Allowing different values for $\alpha$, we repeat the fitting procedure for the quadruple-imaged systems discussed in Sect. \ref{quad}. The result is basically the same as for the HK model, with the parameters listed in Table \ref{table2} not significantly changing. In case of the Jaffe profile ($\alpha=0$), for instance, inclination and PA are altered by about 5$^{\circ}$ and the predicted mass by approximately 10\%.

To further illuminate this obvious insufficiency of our model, let us have a more detailed look at the caustic structure, taking the system PG1115+080 as an example: Choosing a plausible setting for the lens system in MOND, we fix its size to $h=0.72$kpc and the PA to $77.2^{\circ}$ (observed value).
In accordance with the best-fit results, we additionally assume a lens mass of $M=8\times 10^{11}\Msun$ and vary the Dehnen index $\alpha$, the model's ``discyness" $b/(h+b)$ as well as the inclination on a range from $-1$ to $1$, $0.1$ to $0.9$ and $10^{\circ}$ to $90^{\circ}$, respectively. For a selection of such lenses, the corresponding critical curves and caustics are shown in Fig. \ref{selection}. Then, among all resulting lens models, we select those which exhibit the strongest (non-spherical) shear, corresponding to a large astroid caustic size. Since the lens mass should be close to the stellar mass ($M/M_{*}\simeq 1$) in MOND/TeVeS, the idea is now to stepwise decrease the mass of these models. In all cases, we find that, due to the caustics' contraction, the source crosses the astroid caustic way before $M/M_{*}$ reaches unity, thus not corresponding to a quadruple-imaged system anymore. Typically, the crossing seems to take place when the lens model's mass is roughly around $4-6\times 10^{11}\Msun$. For $\alpha=1$, $b/(h+b)=0.38$ and an inclination of $44.5^{\circ}$, this situation is illustrated in the left and middle panel of Fig. \ref{SIS}. Note that we have kept the source position fixed at $(-0.011,0.091)\arcsec$ for our analysis, with the lens being centred at the origin.

Again, this provides a possible explanation why the Dehnen-Kuzmin model (including the HK model) mostly fails to fit quadruple-imaged systems, supporting our earlier conclusion from Sect. \ref{maxshear}. Given that $M/M_{*}\simeq 1$ in MOND, our model is obviously not able to generate sufficiently strong shear (hence large caustics) and a large Einstein ring at the same time. 
For comparison, we also present the resulting caustics and critical curves of a best-fit SIS+$\gamma_{ext}$ model in the right panel of Fig. \ref{SIS}. As is known, its deflection potential can be expressed as
\begin{equation}
\Psi(\xi,\theta)=c\xi+\frac{\gamma}{2}\xi^2{\cos}\left (2(\theta-\theta_{\gamma})\right ).
\label{sisext}
\end{equation}
Choosing the lens' position $(x_l,y_l)=(0.0028,0.0048)\arcsec$, $c=1.14{\arcsec}$, $\gamma=0.07$ and $\theta_{\gamma}=88.7^{\circ}$, the above model is able to fit observations of PG1115+080 satisfyingly.

\section{Summary and discussion}

In this paper, we presented a class of analytic non-spherical models, called the Hernquist-Kuzmin models, that we applied to fit double-imaged and quadruple-imaged lens systems in the context of MOND, using the covariant framework of TeVeS. These analytic models interpolate smoothly between a Hernquist sphere and a Kuzmin disc, and their  great advantage is that the solenoidal field of MOND, which normally appears in Eq. \eqref{eq:A} for non-spherical configurations, is zero. Note that these models are not, stricto sensu, bulge-disc models (see Fig.~\ref{contour}).

After having worked out the lensing properties of the models, we devised a fitting procedure in order to fit 15 double-imaged systems and four quadruple-imaged systems in the CfA-Arizona Space Telescope Lens Survey (CASTLES). Since the fitting problem is ill-posed, especially in the case of double-imaged systems, we used a regularization method to ensure the uniqueness of the solution, by penalizing fits deviating from the Fundamental Plane as well as face-on and discy fits, and fits with an anomalous mass-to-light ratio or a large flux anomaly. 

From this fitting procedure, we find that our model is able to describe the observed image positions of all analyzed double-imaged systems, and we can safely conclude that 10 of these systems yield plausible parameters within the context of MOND/TeVeS. Additionally, our analytic model is able to explain the flux ratios of these binaries in almost every case. One can thus conclude that, in general, galaxy lenses do not need dark matter in MOND/TeVeS. Note that the implied masses for most of these lenses are quite similar to those derived from the spherically symmetric models of \cite{lenstest}, but that the big advantage of our non-spherical model is its ability to fit the precise image-positions rather than just the size of the Einstein ring.

On the other hand, 5 double-imaged systems do not provide a reasonable fit: 
while for two of these systems, the found problems are likely to be solved by considering observational uncertainties, a more accurate model and/or additional effects such as extinction and microlensing, the other three lenses appear to be lacking an obvious explanation\footnote{Note, however, that the stellar mass estimates depend on the adopted initial mass function and star formation rate, and can vary by a factor of 4 in the R-band, which could partly solve the problem of the mass-ratio discrepancy, but not the flux ratio anomalies.}.
It is however quite striking that all these remaining outliers are actually residing in (or close to) {\it groups or clusters} of galaxies. This means that non-linear effects could have an important incidence, since lensing in MOND is much more sensitive to the three-dimensional distribution of the lens (and of the environment) than in GR \citep[e.g.][]{asymmetric}. Moreover, it is known for a while that additional dark matter is needed for clusters of galaxies in MOND \citep[e.g.][]{sanders99,sanders03,neutrinos2,tevesfit,milgromchristmas}, and it has recently been shown that this was the case for groups, too \citep{group}. Possible explanations for this ``cluster dark matter" in the context of MOND range from the presence of numerous clouds of cold gas \citep{milgromchristmas} to the existence of sterile neutrinos with a $5$eV mass \citep{ursaneutrinos}, through the non-trivial effects of the vector field (or of an additional scalar field) in covariant formulations of MOND other than TeVeS \citep[e.g.][]{covariant1,vectornew}. In any case, many studies, including the recent analysis of the velocity dispersions of globular clusters in the halo of NGC~1399 \citep{ngc1399}, have also provided evidence for such dark matter on {\it galaxy scales} in MOND: this is typical for galaxies residing at the centre of clusters only, and can be interpreted \citep{ngc1399} as a small-scale variant of the aforementioned problem of MOND in clusters. This could thus provide an additional, physical, reason to the poor fits obtained here for the two-image lenses residing in groups or clusters.

For the four quadruple-imaged systems, it is a different story: the only acceptable fit is obtained for the Einstein cross Q2237+030, but even in this case, the observed flux ratios cannot be reproduced. However, the anomalous flux ratios here are most likely due to microlensing effects which have not been considered in our analysis. We can thus conclude that MOND does not provide a solution to the flux anomaly issue, mainly because smooth MOND models naturally predict smooth amplification patterns. Among the 3 very poorly fitted lenses, only PG1115+080 and B1422+231 appear in a crowded environment, which could cause the same perturbing effects as for non-isolated double-imaged systems; the remaining lens, SDSS0924+0219, appears relatively isolated. 

We thus argue that, especially in this particular case, the poor fits are due to the intrinsic limitation of our analytic models: we indeed showed that our models were unable to produce a large Einstein ring and a large shear at the same time. Although we did not present a rigorous proof, we tried to make this limitation plausible by analyzing the maximum non-spherical shear of a TeVeS Kuzmin lens as well as the caustic structure of different Hernquist-Kuzmin models. To make the mass more concentrated, we also tried a Dehnen model \citep{dehnen}, but this did not provide a satisfactory solution either. Again, note that our analysis did not consider any contribution due to external shear effects.

We therefore conclude that our analytic models provide good MOND fits to the image positions of isolated two-image lenses, but that some problems are encountered for non-isolated two-image lenses. On the other hand, we showed that our models were barely able to fit quadruple-imaged systems. The present study has thus pinpointed some lenses for which a full three-dimensional numerical model should be devised in MOND. While these analytic models did obviously not yet represent a definitive test of MOND with gravitational lensing, they nevertheless provided a new step in understanding this quite unexplored research area, and in isolating the possibly challenging lensing systems for the future.

\section*{Acknowledgments}

HYS and BF acknowledge hospitality at the University of St Andrews.
HYS thanks Prasenjit Saha and Muhammad Yusaf for valuable comments on an earlier version of the manuscript.
BF is supported by the FNRS, HSZ acknowledges partial support from the Chinese National Science Foundation (Grant
No. 10233040) and a UK PPARC Advanced Fellowship.  
MF is supported by a scholarship from the Scottish Universities Physics Alliance (SUPA).
We thank the referee Bob Sanders for insightful comments and useful suggestions.

\bibliographystyle{mn2e}
\bibliography{quad2}




\end{document}